\documentclass[journal,final,twoside]{IEEEtran}

\usepackage[usenames]{color}
\usepackage{graphicx}
\usepackage{balance}
\usepackage{amsmath}
\usepackage{amssymb}
\usepackage{multirow}
\usepackage{cite}
\usepackage{ifpdf}
\usepackage{epstopdf}
\usepackage{epsf}
\usepackage{hyperref}
\usepackage{fancyhdr}

\usepackage[printonlyused]{acronym}

\usepackage{amsfonts}
\usepackage{amsmath}
\usepackage{mathrsfs}
\usepackage{balance}
\usepackage{cite}
\usepackage{subfigure}

\usepackage{listings}

\usepackage{color}
\usepackage{units}
\usepackage{nicefrac}
\usepackage[printonlyused]{acronym}
\usepackage{pstricks, pst-plot, pst-grad, pstricks-add, pst-node, pstricks-add}
\usepackage[section] {placeins}
\usepackage{fancyhdr}
\definecolor{NewText}{gray}{0}
\newrgbcolor{darkgreen}{0.2 0.6 0.2}
\newrgbcolor{darkred}{0.6 0.2 0.2}
\newrgbcolor{darkblue}{0.2 0.2 0.6}

\acrodef{pdf}[pdf]{probability density function}
\acrodef{cdf}[cdf]{cumulative distribution function}
\acrodef{LT}[LT]{Laplace transform}
\acrodef{BPP}[BPP]{binomial point process}
\acrodef{PPP}[PPP]{Poisson point process}
\acrodef{pgfl}[pgfl]{probability generating functional}

\begin{document}
\abovedisplayskip=0pt
\belowdisplayskip=0pt

\setlength\abovecaptionskip{0.3\baselineskip}
\setlength\belowcaptionskip{-0.25cm}

\hyphenation{multi-symbol}
\title{A Direct Approach to Computing \\ Spatially Averaged Outage Probability}
\author{ Matthew~C.~Valenti,~\IEEEmembership{Senior~Member,~IEEE,}
  Don~Torrieri,~\IEEEmembership{Senior~Member,~IEEE,} \\ and
  Salvatore~Talarico~\IEEEmembership{Student Member,~IEEE}
  \thanks{Manuscript received Feb. 18, 2014; revised Apr. 3, 2014.  The associate editor coordinating the review of this letter and approving it for publication was M. Di Renzo.}
  \thanks{M.~C.~Valenti and S.~Talarico are with West Virginia University, Morgantown, WV (email:valenti@ieee.org, Salvatore.Talarico81@gmail.com).}
  \thanks{D. ~Torrieri is with the US Army Research Laboratory, Adelphi, MD (email: don.j.torrieri.civ@mail.mil).}
  \thanks{Digital Object Identifier 10.1109/LCOMM.2014.XXXXXX.XXXXXX}
  \vspace{-0.05cm}
}

\date{}
\maketitle

\begin{abstract}
This letter \textcolor{NewText}{describes a direct method} for computing the spatially averaged outage probability of a network with interferers located according to a point process and signals subject to fading.  \textcolor{NewText}{Unlike most common approaches, it} does not require transforms such as a Laplace transform.  \textcolor{NewText}{Examples show how to directly obtain the outage probability in the presence of Rayleigh fading in networks whose interferers are drawn from}  binomial and Poisson point processes defined over arbitrary regions.  We furthermore show that, by extending the arbitrary region to the entire plane, the result \textcolor{NewText}{for Poisson point processes} converges to the same expression found by Baccelli et al..
\end{abstract}

\begin{IEEEkeywords}
Outage probability, stochastic geometry, point processes, interference modeling, fading.
\end{IEEEkeywords}

\IEEEpubid{XXXX-XXXX/XX\$31.00~\copyright~2014 IEEE}

\IEEEpubidadjcol


\vspace{-0.4 cm}
\section{Introduction} \label{Section:Intro}
\IEEEPARstart{T}{he}
spatially averaged outage probability is a useful and popular metric for characterizing the performance of wireless networks, as it captures in a single quantity the dynamics of both the channel (e.g., fading and shadowing) and the random locations of the interferers. The common approach to computing the spatially averaged outage probability is to assume that the network configuration is modeled as a spatial point process, such as a \ac{BPP} or a \ac{PPP}, and then use tools from {\em stochastic geometry}
\cite{haenggi:2012,baccelli:2006,
ganti:2009b,
andrews:2010,elsawy:2013}
to compute the outage probability.

As outlined in \cite{elsawy:2013}, there are five main techniques used in the literature for obtaining the spatially averaged outage probability.  The first technique assumes that the reference link undergoes Rayleigh fading and obtains the outage probability in the form:
\begin{eqnarray}
\epsilon
& = &
1
-
\exp \left\{
- \frac{\beta}{\mathsf{SNR}}
\right\}
{\mathcal L}_I(s) \Big|_{s=c\beta}
\end{eqnarray}
where ${\mathcal L}_I(s)$ is the \textcolor{NewText}{\ac{LT} of the \ac{pdf}} of the aggregate interference $I$, $\beta$ is the outage threshold, $\mathsf{SNR}$ is the average signal-to-noise ratio, and $c$ is a constant. The key \textcolor{NewText}{is to find the \ac{LT} of the \ac{pdf} of $I$}, which thanks to the \ac{pgfl}  is easy to compute for many cases involving a \ac{PPP}.
The second technique is to only consider the most dominant interferer or the nearest $n$ neighbors, which can be used to obtain a bound on outage probability.  The third approach is to use simulations to empirically fit the \ac{pdf} of $I$ to a known distribution, such as a Gaussian or shifted lognormal.    The fourth approach involves the use of the Plancherel-Parseval theorem, which replaces the need for inverting the \ac{LT} with a complicated integral. The fifth approach is to invert the \ac{LT} numerically.

Finding the outage probability of an arbitrarily shaped network with interferers drawn from a \ac{BPP}\textcolor{NewText}{, which we refer to as the \emph{\ac{BPP} outage probability},} is a nontrivial problem that has been considered in the literature by using the aforementioned existing methods \cite{srinivasa:2007,srinivasa:2010,guo:2013}.
\textcolor{NewText}{Another method for obtaining the BPP outage probability is suggested by \cite{torrieri:2012} and \cite{guo:2013}.  A key feature of this approach}
is that, unlike the other techniques, transforms such as the \ac{LT} are not used or needed.
\textcolor{NewText}{In \cite{guo:2013}, the outage probability is first conditioned on $I$ and expressed in terms of the \ac{cdf} of $g_0$, which is the fading gain of the reference link.  The conditioning on $I$ is removed by first removing the conditioning on the $\{g_1, ..., g_M\}$, which are the fading gains of the $M$ interferers, and then removing the conditioning on $\{r_1,...,r_M\}$, which are the distances to the interferers.  In \cite{torrieri:2012}, the outage probability is conditioned on the $\{r_i\}$, which involves a marginalization over the gains of the reference and interfering links, $\{g_0, ..., g_M\}$.  The conditioning on $\{r_i\}$ is then removed, but closed-form expressions are found only for the case that the interferers are distributed on a ring or disk.}

\textcolor{NewText}{This paper reviews the steps of \cite{torrieri:2012} for obtaining the \ac{BPP} outage probability, focusing on the case of Rayleigh fading.  As in \cite{guo:2013}, the result for an arbitrary topology is expressed as a one-dimensional integral.   Like \cite{guo:2013}, we present results for the case that the network is shaped as a regular polygon and the reference receiver is at its center, but our results are expressed as an infinite sum.  We then present an effective approach for handling arbitrary topologies by evaluating the one-dimensional integral either numerically or by Monte Carlo simulation, which is the first main contribution of the paper.}

\IEEEpubidadjcol

\textcolor{NewText}{Next, as the second main contribution, we extend the direct approach to the problem of finding the \ac{PPP} outage probability.
In particular, we}
show how to obtain the outage probability without using the \ac{pgfl} when the interferers are drawn from a PPP defined over the entire plane.
First, the interferers are restricted to a finite region, and the PPP outage probability is found by averaging the BPP outage probability with respect to the number of interferers in the region, which for a PPP will be a Poisson variable.  Next, the boundaries of the restricted area are allowed to go to infinity.
The resulting outage probability expression exactly coincides with the one in the seminal reference by Baccelli et al. \cite{baccelli:2006}.

\section{Direct Approach to Spatial Averaging}
\label{Section:SystemModel}
Consider a network comprising a reference receiver, a reference transmitter $X_{0}$, and $M$ interfering transmitters $X_{1},...,X_{M}.$    The coordinate system is chosen such that the reference receiver is located at the origin, and distance is normalized such that the reference transmitter is at unit distance from the receiver.  The interferers are located within an arbitrary two-dimensional region $\mathcal A$, which has area $|\mathcal A|$.   The number of interferers within $\mathcal A$ could be fixed
or random.
Let $r_i$ denote the distance from $X_i$ to the receiver, and let $\boldsymbol{r }=\{r_{1},...,r _{M}\}$ represent the set of distances to the interferers,
which
corresponds to a specific network topology.

With probability $p_i$, $X_{i}$ transmits a signal with power $P_i$.
Though not a requirement for the analysis, we assume here that $p_i = p$ and $P_i = P_0$ for all $i \in \{1,...,M\}.$
The instantaneous signal-to-interference-and-noise ratio (SINR) at the receiver is
\begin{eqnarray}
   \gamma
   & = &
   \frac{ g_0  }{ \displaystyle \mathsf{SNR}^{-1} + \sum_{i=1}^M I_i g_i r_i^{-\alpha} }\label{Equation:SINR2}
\end{eqnarray}
where
$g_i$ is the power gain due to fading, $\alpha > 2$ is the attenuation power-law exponent,
and the $\{I_i\}$ are independent and identically distributed (i.i.d.) Bernoulli variables with $\mathbb P[I_i=1] = p$.

An {\em outage} occurs when the instantaneous SINR falls below a threshold $\beta$.   Let $\epsilon(\mathbf{r})$ represent the outage probability
\textcolor{NewText}{at the reference receiver} in the
topology associated with $\boldsymbol{r }$,
\begin{eqnarray}
   \epsilon(\mathbf{r})
   & = &
   \mathbb P \left[ \gamma \leq \beta \big| \boldsymbol r \right]\hspace{-0.1cm}.
   \label{Equation:Outage1}
\end{eqnarray}

When all transmissions are subject to Rayleigh fading, the $\{g_i\}$ are i.i.d. unit-mean exponential random variables, and
\begin{eqnarray}
\epsilon(\mathbf{r})
& = &
1 - e^{-\beta/\mathsf{SNR}} \prod_{i=1}^M \left(  1 - p + \frac{p r_i^{\alpha} }{\beta + r_i^{\alpha} } \right)\hspace{-0.1cm}.
\label{Equation:RayleighConditional}
\end{eqnarray}
The above expression can be obtained from \cite{torrieri:2012} by substituting its (30) into its (12).  Alternatively, it can be found from \cite{ganti:2009b} by substituting its (2.3) into its (1.4).

The \emph{spatially averaged} outage probability is found by taking the expectation of $\epsilon(\mathbf{r})$ with respect to the network geometry.
Let $\epsilon[M]$ denote the spatially averaged outage probability of a network with a fixed number of interferers, which is found by taking the expectation of $\epsilon(\mathbf{r})$ with respect to $\mathbf{r}$ under the condition that there are $M$ interferers; i.e.,
\begin{eqnarray}
   \epsilon[M] \hspace{-0.1cm}
   & = &
     \mathbb E_{\mathbf{r}} \left[ \epsilon(\mathbf{r}) | M \right]
   =
   \int
   \epsilon(\mathbf{r})
   {f}_\mathbf{r}\left( \mathbf{r} | M \right)
    d\mathbf{r}
    \label{epsilon_M}
\end{eqnarray}
where ${f}_\mathbf{r}\left( \mathbf{r} | M \right) $ is the \ac{pdf} of $\mathbf{r}$
when there are $M$ interferers.

When the number of interferers is random, then the overall spatially averaged outage probability, which we denote $\epsilon$, can be found by taking the expectation of $\epsilon[M]$ with respect to the number of interferers in the region\footnote{It is noted that, for the specific case of a PPP,
\cite{sousa:1990} also conditions on $M$ and takes
the expected value of $\epsilon[M]$ with respect to $M$,
but \cite{sousa:1990} does not consider fading and operates in a transformed domain,
which is not necessary with the present approach.}
. In particular,
  \begin{eqnarray}
  \epsilon
  & = &
  \mathbb E_M \left[ \epsilon [M] \right]
   =
   \sum_{ m=0}^{ \infty } p_{M}[m] \epsilon[m]
   \label{cdf_PPP}
\end{eqnarray}
where $p_{M}[m]$ is the probability mass function (pmf) of $M$.

\section{Binomial Point Processes}
\label{Section:BPP}
Assume that the $M$ interferers are independently and uniformly distributed (i.u.d.) over $\mathcal A$.
Thus, the interferers are drawn from a BPP of intensity $\lambda = M/|\mathcal{A}|$ \cite{haenggi:2012}. Because the $\{r_i\}$ are independent,
\begin{eqnarray}
   {f}_\mathbf{r}\left( \mathbf{r} | M \right)
   & = &
   \prod_{i=1}^M
   f_{r}(r_i).
   \label{indpdf}
\end{eqnarray}
where $f_{r}(r_i)$ is the \ac{pdf} of $r_i$.

In Rayleigh fading, the spatially averaged outage probability can be found by substituting (\ref{Equation:RayleighConditional}) and (\ref{indpdf})  into (\ref{epsilon_M}) and using the fact that the $\{r_i\}$ are identically distributed:
\begin{eqnarray}
\epsilon[M]
\hspace{-0.15cm}
& = &
\hspace{-0.15cm}
1 - e^{-\beta/\mathsf{SNR}}
\left(  1 - p +
p \int
f_r(r)
\frac{r^{\alpha} }{\beta + r^{\alpha} } dr \right)^M\hspace{-0.35cm}.
\label{cdfwithintegral}
\end{eqnarray}
It follows that computing the outage probability for a BPP boils down to evaluating a single one-dimensional integral.
The following examples show how to obtain the outage probability for regular and arbitrary shapes.

{\bf Example \#1}.  Assume that $\mathcal A$ is an L-sided regular polygon inscribed in a circle of radius $r_{\mathsf{out}}$. The polygon is centered at the origin, and a circular {\em exclusion zone} of radius $r_\mathsf{in}$ is removed from the center of the polygon to ensure that all interferers are at least distance $r_\mathsf{in}$ away.  The  area of this region is $|\mathcal{A}| = \frac{1}{2} L r_{\mathsf{out}}^2 \sin\left( \frac{2 \pi}{L}\right) - \pi r_{\mathsf{in}}^2$.  Let $r_\mathsf{c} = r_\mathsf{out} \sin ( \pi(L-2)/(2L) )$ be the distance from the origin to any corner of the polygon.  From \cite{khalid:2013}, the identically distributed $\{r_i\}$ have \ac{pdf}
 \begin{eqnarray}
   {f}_{r_i}( r)
   =
   \begin{cases}
\frac{2 \pi r }{|{\mathcal A}|}
      &
      \mbox{\small for $ r_{\mathsf{in}} \leq r \leq r_\mathsf{c}$} \\
\frac{2 \pi r }{|{\mathcal A}|}
 -
 \frac{2 L r }{|{\mathcal A}|}
\arccos \left(  \frac{ r_\mathsf{c} }{ r } \right)
      &
       \mbox{\small for $  r_\mathsf{c} \leq  r \leq  r_{\mathsf{out}} $}
   \end{cases} \label{pdf_r_Lpolygon}
\end{eqnarray}
and zero elsewhere.

When (\ref{pdf_r_Lpolygon}) is substituted into  (\ref{cdfwithintegral}), the integral is
\begin{eqnarray}
\int f_r(r) \frac{r^{\alpha} }{\beta + r^{\alpha} } dr
& = &
T_1 - T_2
\label{t_1_minus_t_2}
\end{eqnarray}
where
\begin{eqnarray}
T_1
& = &
\frac{2 \pi }{|\mathcal{A}|}
\int_{ r_\mathsf{in} }^{ r_\mathsf{out} }
\frac{r^{\alpha+1} }{\beta + r^{\alpha} } dr \\
T_2
& = &
\frac{2  L }{|\mathcal{A}|} \int_{r_{\mathsf{c}}  }^{ r_\mathsf{out} }
\frac{r^{\alpha+1}}{\beta + r^{\alpha} } \arccos \left( \frac{r_{\mathsf{c}} }{r}\right) dr.
\label{integral_T12}
\end{eqnarray}
The first integral is
\begin{eqnarray}
\displaystyle
T_1
\hspace{-0.25cm} & = & \hspace{-0.25cm}
\frac{ 2 \pi }{ \beta |\mathcal{A}|}
\left[
\Psi_1 \left( r_\mathsf{out} \right)
-
\Psi_1 \left( r_\mathsf{in}  \right)
\right] \label{split}
\end{eqnarray}
where
\begin{eqnarray}
\Psi_k(y)
& = & \
\int_{ 0 }^{ y }
\frac{ r^{\alpha+k} }{1 + \frac{1}{\beta} r^{\alpha} } dr.
\end{eqnarray}
By performing the change of variable $\nu = (r/y)^\alpha \hspace{-0.18cm}$,
\begin{eqnarray}
\hspace{-0.3cm} \Psi_k(y)
&\hspace{-0.3cm} = & \hspace{-0.3cm}
\frac{y^{\alpha+k+1}}{\alpha}
\int_{ 0 }^{ 1 }
\nu^{\frac{1+k}{\alpha}}
\left( 1 +
\frac{y^\alpha}{\beta} \nu
\right)^{-1}
d \nu \nonumber \\
\hspace{-1.0cm}
& \hspace{-1.2cm}=  & \hspace{-0.6cm}
\left(\hspace{-0.05cm}
\frac{y^{\alpha+k+1}}{1\hspace{-0.05cm}+\hspace{-0.05cm}k\hspace{-0.05cm}+\hspace{-0.05cm}\alpha}
\hspace{-0.05cm}\right)
\hspace{-0.1cm}_{2}F_{1} \hspace{-0.1cm}
\left(\hspace{-0.05cm}
\left[ 1, \hspace{-0.05cm}\frac{1\hspace{-0.05cm}+\hspace{-0.05cm}k}{\alpha} \hspace{-0.05cm}+ \hspace{-0.05cm}1 \right]\hspace{-0.05cm};\hspace{-0.05cm}
\frac{1\hspace{-0.05cm}+\hspace{-0.05cm}k}{\alpha}\hspace{-0.05cm} + \hspace{-0.05cm} 2;
- \frac{y^\alpha}{\beta}\hspace{-0.05cm}
\right) \label{Ihyp}
\end{eqnarray}
where
$_{2}F_{1}$ is  the Gauss hypergeometric function,
\begin{eqnarray}
\hspace{-0.4cm}_{2}F_{1}([a,b];c;x) \hspace{-0.2cm} &=& \hspace{-0.2cm} \nonumber \\
 \hspace{-0.4cm} && \hspace{-2.3cm} \frac{\Gamma (c)}{\Gamma (b)\Gamma (c-b)}%
\int_{0}^{1}\nu ^{b-1}(1-\nu )^{c-b-1}(1-x\nu )^{-a}d\nu  \label{HYPERG}
\end{eqnarray}%
and $\Gamma(\cdot)$ is the gamma function.  Note that (\ref{HYPERG}) does not converge when $c$ is a non-positive integer.

\textcolor{NewText}{The above analysis agrees with \cite{guo:2013}, which does not include an exclusion zone ($r_\mathsf{in}=0$),
is generalized to Nakagami-m fading, and uses numerical integration to evaluate the integral. Alternatively, the}
second integral can be found by substituting the Taylor series expansion of $\arccos$ into (\ref{integral_T12}),
resulting in
\begin{eqnarray}
T_2
& = &
\frac{ 2  L }{ \beta | \mathcal A|}
\left[
K \left( r_\mathsf{out} \right)
-
K \left( r_\mathsf{c} \right)
\right]
\label{eqn:t_2}
\end{eqnarray}
where
\begin{eqnarray}
K( y )\hspace{-0.05cm}
=\hspace{-0.05cm}
\frac{\pi}{2}
\Psi_1(y)
\hspace{-0.05cm}- \hspace{-0.05cm}\sum_{ n=0 }^{ \infty }\hspace{-0.05cm}
\frac{ \left( 2n\right)! \hspace{-0.05cm} \left(\hspace{-0.05cm} r_{\mathsf{c}}
\hspace{-0.05cm} \right)^{2n+1}}{4^n \left( n!\right)^2 \left( 2n+1\right)}\hspace{-0.05cm}
\Psi_{-2n}(y).
\label{K}
\label{T2_1}
\end{eqnarray}
By substituting (\ref{t_1_minus_t_2}) into (\ref{cdfwithintegral}) with $T_1$ given by (\ref{split}) and $T_2$ given by (\ref{eqn:t_2}),
the spatially averaged outage probability is
\begin{eqnarray}
\epsilon[M] \hspace{-0.05cm}
=\hspace{-0.05cm}
1 \hspace{-0.05cm} - \hspace{-0.05cm} e^{-\beta / \mathsf{SNR}}
\left(
1 + \frac{p}{| \mathcal A |}
\Theta( r_\mathsf{in}, r_\mathsf{c}, r_\mathsf{out} )
\right)^M
\label{cdf_BPP_Polygons}
\end{eqnarray}
where
\begin{eqnarray}
  \hspace{-0.4cm} \Theta( x, y, z )
  \hspace{-0.3cm}
  & = &
  \hspace{-0.3cm}
  \pi \left[
  \Phi( z ) \hspace{-0.05cm} - \hspace{-0.05cm} \Phi( x )
  \right]
  \hspace{-0.05cm} -\hspace{-0.05cm}
  \frac{2L}{\beta}
  \left[
  K( z ) \hspace{-0.05cm} - \hspace{-0.1cm} K( y )
  \right]\hspace{-0.1cm} - \hspace{-0.1cm} C(z) \\
\hspace{-0.15cm} \Phi( x )
 \hspace{-0.1cm}
 & = & \hspace{-0.1cm}
 \frac{2 }{\beta }
 \Psi_1(x)
 - x^2
\label{Psi} \\
\hspace{-0.15cm} C (x)
 \hspace{-0.1cm}
 & = & \hspace{-0.1cm}
 x^2\left[ \frac{1}{2} L \sin\left( \frac{2 \pi}{L} \right) -\pi \right].
\label{C}
\end{eqnarray}

{\bf Example \#2}.  Assume that the interferers are in an annular region with inner radius
$r_\mathsf{in}$ and outer radius $r_\mathsf{out}$.
Clearly, this shape is the limiting case of the polygon given in Example \#1 as $L \rightarrow \infty$.  In this case, $r_\mathsf{c}\rightarrow r_\mathsf{out}$, $C(r_\mathsf{out}) \rightarrow 0$, and  (\ref{cdf_BPP_Polygons}) becomes
\begin{eqnarray}
\epsilon[M]
& = &
1 - e^{-\beta / \mathsf{SNR}}
\left(
1 +
\frac{p \pi }{ |\mathcal{A}| }
\left[
\Phi \left( r_{\mathsf{out}} \right)
-
 \Phi \left( r_{\mathsf{in}} \right)
\right]
\right)^M
\label{cdf_BPP}
\end{eqnarray}
where $|\mathcal A| = \pi( r_{\mathsf{out}}^2 - r_{\mathsf{in}}^2 )$.

{\bf Example \#3}. Assume that the interferers are in an arbitrarily shaped area. In this case, the integral in (\ref{cdfwithintegral}) cannot generally be evaluated in closed form.  However, it is just a one-dimensional integral and as such, it can easily be evaluated numerically.  The integral can be expressed as
\begin{eqnarray}
\int
f_r(r)
\frac{r^{\alpha} }{\beta + r^{\alpha} } dr
& = &
\mathbb E_r \left[ \frac{r^{\alpha} }{\beta + r^{\alpha} } \right].
\label{arbitrary}
\end{eqnarray}
The right side of (\ref{arbitrary}) suggests a simple numerical approach to solving the integral: Draw a large number of points $\{r_i\}$ distributed according to $f_r(r)$, and for each one of the points evaluate the function  $g(r) = (1+\beta r^{-\alpha})^{-1}$.  The integral is then well approximated by the average of the $\{ g(r_i) \}$.
\textcolor{NewText}{This is an alternative to \cite{guo:2013}, which requires $f_r(r)$ to be determined and substituted into the integral before it is numerically integrated.}
When the process is homogeneous, the $\{r_i\}$ can be selected by placing the shape inside a box, randomly selecting a point with uniform probability in the box and if the point falls within the shape, including it in the set of $\{r_i\}$, then repeating the process until $\{r_i\}$ is sufficiently large.  This approach is a Monte Carlo simulation, but it is a simple simulation of the location of a single interferer.  Importantly, it is not a simulation that requires the fading coefficients to be realized, and it does not require all $M$ interferers to be placed.  Alternatively, the $\{r_i\}$ could be selected by overlaying the shape with a fine grid of equally spaced points.

\section{Poisson Point Processes}
\label{Section:PPP}
Suppose that the interferers are drawn from a PPP with intensity $\lambda$ on the plane $\mathbb R^2$.  Let $\epsilon(\mathcal A)$ denote the outage probability considering only those interferers located within region $\mathcal A$.  The number of interferers $M$ within region $\mathcal A$ is Poisson with mean $\mathbb E[M] = \lambda |\mathcal A|$.  It follows that the pmf of the number of interferers within $\mathcal A$ is given by \cite{haenggi:2012}
\begin{eqnarray}
p_{M}[m]
& = & \frac{(\lambda |\mathcal A|)^{m}}{m!}e^{-\lambda |\mathcal A|}, \; \; \; \mbox{for $m \geq 0$}.
\label{pmf_Poisson}
\end{eqnarray}
The outage probability can then be found for any arbitrary $\mathcal A$ by substituting (\ref{pmf_Poisson}) and the appropriate $\epsilon[m]$ into ($\ref{cdf_PPP}$).


{\bf Example \#4}.  Suppose that $\mathcal A$ is the polygon described in Example \#1.   Substituting ($\ref{pmf_Poisson}$) and ($\ref{cdf_BPP_Polygons}$) into ($\ref{cdf_PPP}$) gives
\begin{eqnarray}
\epsilon( \mathcal A)
\hspace{-0.25cm}
& = &
\hspace{-0.25cm}
1 - e^{-\lambda |\mathcal{A}| - \beta  /\mathsf{SNR}}
\hspace{-0.1cm}
\sum_{ m=0}^{ \infty }
\hspace{-0.05cm}
\frac{1}{m!}
\hspace{-0.05cm}
\left(
\lambda |\mathcal{A}|
\hspace{-0.05cm} + \hspace{-0.05cm}
\lambda p
\Theta( r_\mathsf{in}, r_\mathsf{c}, r_\mathsf{out} )
\right)^m \hspace{-0.25cm}. \nonumber
\end{eqnarray}
From the power-series representation of the exponential function, we find that the outage probability is found to be
\begin{eqnarray}
\epsilon( \mathcal A )
\hspace{-0.25cm}
& = &
\hspace{-0.25cm}
1 -
\exp \left\{ -\frac{\beta}{\mathsf{SNR}}   + \lambda p
\Theta( r_\mathsf{in}, r_\mathsf{c}, r_\mathsf{out} )
\right\}.
\label{PPP-poly}
\end{eqnarray}

{\bf Example \#5}.  If $\mathcal A$ is a disk of radius $r_\mathsf{out}$, the outage probability can be found from (\ref{PPP-poly}) with $r_\mathsf{in} = 0$ and $r_\mathsf{c} = r_\mathsf{out}$,
\begin{eqnarray}
\epsilon( \mathcal A )
\hspace{-0.25cm}
& = &
\hspace{-0.25cm}
1 -
\exp \left\{ -\frac{\beta}{\mathsf{SNR}}   +
\pi \lambda p
\Phi( r_\mathsf{out} )
\right\}.
\label{PPP-circle}
\end{eqnarray}

{\bf Example \#6}.  Now let $\mathcal A$ include the interferers on the entire plane $\mathbb R^2$.  The outage probability can be found by taking the limit of (\ref{PPP-circle}) as $r_\mathsf{out} \rightarrow \infty$.  The limit may be expressed as
\begin{eqnarray}
\hspace{-0.5cm}
\epsilon
\hspace{-0.25cm}
& = &
\hspace{-0.25cm}
1 -
\exp \left\{ -\frac{\beta}{\mathsf{SNR}}   + \pi \lambda p  \lim_{x \rightarrow \infty}  \Phi \left( x \right)  \right\}.
\label{PPPNN_lim1}
\end{eqnarray}
where the right hand side follows from the continuity of the exponential function.

By performing the change of variables $y=\frac{\beta}{x^{\alpha}}$, the limit in (\ref{PPPNN_lim1}) may be written as
\begin{eqnarray}
\hspace{-0.1cm}
 \lim_{x \rightarrow \infty}  \Phi \left( x \right)
 \hspace{-0.25cm}
 & = &
 \hspace{-0.25cm}
 \lim_{y \rightarrow 0}  \Phi\left( \frac{ \beta^{1/\alpha} }{ y^{1/\alpha} } \right)
=
  \frac{2}{\alpha}\beta^{\frac{2}{\alpha}} \lim_{y \rightarrow 0} \left\{ y^{-\frac{2}{\alpha}} \phi \left( y \right) \right\} \nonumber \\
\label{Psi3}
\end{eqnarray}
\vspace{-0.65cm}

\noindent where
\begin{eqnarray}
\phi(y)
& \hspace{-0.25cm} = \hspace{-0.25cm} &
\frac{1}{y} \int_{0}^{1} t^{\frac{2}{\alpha}} \left(1+\frac{t}{y} \right)^{-1} dt
- \frac{\alpha}{2}.
\label{Psi6}
\end{eqnarray}
By splitting the integral at $y$ and factoring the second integrand,
\begin{eqnarray}
\phi(y)
& \hspace{-0.25cm} = \hspace{-0.25cm} &
\frac{1}{y} \int_{0}^{y} t^{\frac{2}{\alpha}} \left(1+\frac{t}{y} \right)^{-1} dt  \nonumber \\
& & \hspace{-0.5cm} +  \int_{y}^{1} t^{\frac{2}{\alpha}-1} \left(1+\frac{y}{t} \right)^{-1} dt
- \frac{\alpha}{2}.
\label{Psi8}
\end{eqnarray}
By substituting the Taylor series of each of the two integrals and then
integrating term by term,
\begin{eqnarray}
\phi(y)
 \hspace{-0.1cm} = \hspace{-0.1cm}
\frac{1}{y} \sum_{i=0}^{\infty}
\frac{(-1)^{i}y^{\frac{2}{\alpha}+1} }{\frac{2}{\alpha}+1+i}
  \hspace{-0.1cm} +
 \hspace{-0.1cm}
 \sum_{i=0}^{\infty}  \frac{(-1)^{i}  \left( y^i- y^{\frac{2}{\alpha}} \right)}{\frac{2}{\alpha}-i} \hspace{-0.1cm}
  - \hspace{-0.1cm}\frac{\alpha}{2}.
\label{Psi9}
\end{eqnarray}
Absorbing $y$ into the first series and representing the second series as two series,
\begin{eqnarray}
\phi(y)
 \hspace{-0.1cm} = \hspace{-0.1cm}
\sum_{i=0}^{\infty}  \frac{(-1)^{i} y^{\frac{2}{\alpha}} }{\frac{2}{\alpha}+1+i}
+ \sum_{i=0}^{\infty}   \frac{(-1)^{i}y^i}{\frac{2}{\alpha}-i} \nonumber
- \sum_{i=0}^{\infty}   \frac{(-1)^{i}y^{\frac{2}{\alpha}}}{\frac{2}{\alpha}-i}
- \frac{\alpha}{2}.
\label{Psi10}
\end{eqnarray}
Canceling the first term of the second series by the trailing $\alpha/2$ and multiplying both sides by $y^{-\frac{2}{\alpha}}$,
\begin{eqnarray}
 \hspace{-0.5cm}
 y^{-\frac{2}{\alpha}} \phi(y)
\hspace{-0.35cm} &   =    & \hspace{-0.35cm}
 \sum_{i=0}^{\infty}  \hspace{-0.1cm}
 \frac{(-1)^{i}}{\frac{2}{\alpha}+1+i}
 \hspace{-0.05cm}
 +   \hspace{-0.1cm} \sum_{i=1}^{\infty} \hspace{-0.1cm} \frac{(-1)^{i}y^{i-\frac{2}{\alpha}}}{\frac{2}{\alpha}-i}
 -  \hspace{-0.1cm}
  \sum_{i=0}^{\infty}
  \frac{(-1)^{i}}{\frac{2}{\alpha}-i}.
\label{Psi12}
\end{eqnarray}
Taking the limit of (\ref{Psi12}) as $y \rightarrow 0$, the first and last series are unaffected because they are independent of $y$.
Since $\alpha > 2$,
the  exponent of $y^{i-\frac{2}{\alpha}}$ is always positive,
and hence the limit of the second series is zero.  Therefore,
\begin{eqnarray}
\lim_{y \rightarrow 0}
y^{-\frac{2}{\alpha}}  \phi(y)
& = &
 \sum_{i=0}^{\infty}  \hspace{-0.1cm}
 \frac{(-1)^{i}}{\frac{2}{\alpha}+1+i}  \hspace{-0.1cm}
-
  \sum_{i=0}^{\infty}
  \frac{(-1)^{i}}{\frac{2}{\alpha}-i}.
\label{Psi13}
\end{eqnarray}
By performing the change of variables $i=-j-1$,
 \begin{eqnarray}
\hspace{-0.3cm}
\sum_{i=0}^{\infty}   \frac{(-1)^{i}}{\frac{2}{\alpha}+1+i}
\hspace{-0.3cm}
& = &
\hspace{-0.3cm}
\sum_{j=-\infty}^{-1}  \frac{(-1)^{-j-1}}{\frac{2}{\alpha}-j}
\Rightarrow -\sum_{i=-\infty}^{-1}  \frac{(-1)^{i}}{\frac{2}{\alpha}-i}. \label{sum1}
\end{eqnarray}
Substituting (\ref{sum1}) into (\ref{Psi13}) and combining the two series,
\begin{eqnarray}
 \lim_{y \rightarrow 0}
y^{-\frac{2}{\alpha}}  \phi(y)
 \hspace{-0.1cm} & = & \hspace{-0.1cm}
 - \sum_{i=-\infty}^{\infty} \frac{(-1)^{i}}{\frac{2}{\alpha}-i}
 =
 - \pi \csc \left( \frac{2 \pi}{\alpha} \right)
\label{Psi17}
\end{eqnarray}
where the last step follows because the series gives a partial fraction expansion of an
analytic function of $2/\alpha$ that has poles at all integer values
and, hence, is proportional to $\csc(2 \pi/\alpha)$; see also (4.22.5) of \cite{olver:2010}.
Substituting (\ref{Psi17}) into (\ref{Psi3}),
 \begin{eqnarray}
 \lim_{x \rightarrow \infty}  \Phi \left( x \right) =  -\frac{ 2 \pi }{\alpha} \beta^{2/\alpha} \csc \left( \frac{2 \pi}{\alpha} \right).
\label{Psi19}
\end{eqnarray}
Substituting (\ref{Psi19}) into  (\ref{PPPNN_lim1}) yields
\begin{eqnarray}
\epsilon
& = &
1-
\exp \left\{
-\frac{\beta}{\mathsf{SNR}}
- \frac{2 \pi^2 \lambda p}{{\alpha}}\beta^{\frac{2}{\alpha}}
\csc \left(\frac{2\pi}{\alpha}\right) \right\}
\label{Baccelli1}
\end{eqnarray}
which, in the absence of noise ($\mathsf{SNR} \rightarrow \infty$), corresponds to equation (3.4) in \cite{baccelli:2006}.
and equation (61) in \cite{weber:2010}.

\begin{figure}[!t]
\centering
\includegraphics[width=8.9cm]{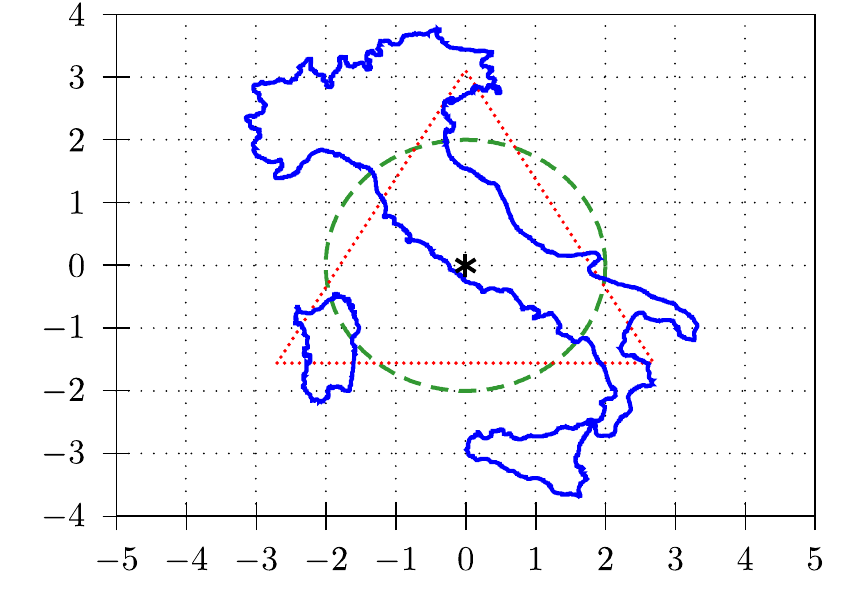}
\vspace{-0.75cm}
\caption{The three network regions considered. \textcolor{NewText}{The location of the reference receiver is indicated by ($\ast$).} \label{Figure:Arenas} }
\vspace{-0.5cm}
\end{figure}

\section{Numerical Results}
\label{Section:Example}
\begin{figure}[!t]
\centering
\includegraphics[width=8.9cm]{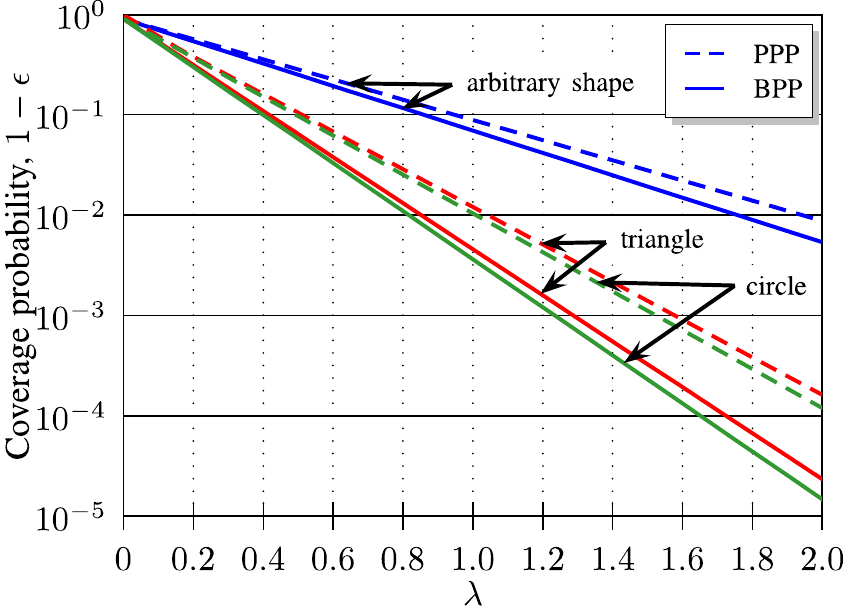}
\vspace{-0.75cm}
\caption{ Coverage probability $\bar{\epsilon} = 1 - \epsilon$ as function of $\lambda$.  Results are shown for the cases that the interferers are drawn from a BPP (solid lines) and from a PPP (dashed lines).  From highest to lowest coverage probability, the curves correspond to the arbitrary shape (blue lines), the triangle (red lines), and the circle (green lines).
\label{Figure:Outage} }
\vspace{-0.5cm}
\end{figure}

We compare the \textcolor{NewText}{outage probability} when the reference receiver is at the center of the three different shapes that are shown in Fig. \ref{Figure:Arenas}.  The shapes are a disk of radius $r_\mathsf{out} = 2$, a triangle ($L=3$ polygon), and an arbitrary shape, which happens to look like a map of Italy including the islands of Sicily and Sardinia (emphasizing that $\mathcal A$ need not be contiguous). In all cases, the shapes are scaled so that they have the same area, $|\mathcal A | =4\pi$, and there is no exclusion zone around the receiver; i.e., $r_\mathsf{in} = 0$.   The spatially averaged outage probability for each network was computed for  $\lambda \leq 2$, contention probability $p=1$, path-loss exponent $\alpha = 3.2$, threshold $\beta = 0$ dB, and average $\mathsf{SNR} = 10$ dB. Results are \textcolor{NewText}{found} for both the BPP and the PPP cases. For the triangle, the series in (\ref{T2_1})
is truncated at the $25^{th}$ term.  For the arbitrary region, (\ref{arbitrary})
is evaluated using a fine grid of $125,640$ equally spaced points.
The coverage probabilities, defined as $\bar{\epsilon} = 1 - \epsilon$ are shown in Fig. \ref{Figure:Outage}.  As can be seen, the disk-shaped network has the worst coverage probability, while those that depart from a circle have better performance.  The \textcolor{NewText}{PPP} coverage probability is better than that of the BPP.  The irregularly shaped network has a very high coverage probability, suggesting that the common assumption of a disk shaped network is overly pessimistic for realistic networks with highly irregular shapes.

\section{Conclusion} \label{Section:Conclusion}

\textcolor{NewText}{While the standard way to find the PPP outage probability is to use the \ac{LT} and \ac{pgfl}, it is possible to obtain the result in a more direct manner.   The starting point is to find the outage probability conditioned on the network topology.  Next, the number of interferers is fixed and the outage probability is averaged with respect their random positions. The result is the BPP outage probability.  Then, the BPP outage probability is marginalized with respect to the number of interferers to obtain the PPP outage probability.  By allowing the network to extend over the entire plane, the classic result of Baccelli et al. is obtained.}



\bibliographystyle{ieeetr}
\bibliography{BaccelliProofRef}

\end{document}